\documentclass{midl} % Include author names
\usepackage{mwe} % to get dummy images
\usepackage{floatrow}
\newfloatcommand{capbtabbox}{table}[][\FBwidth]
\usepackage{caption}
\jmlrworkshop{Full Paper – MIDL 2020 Track}
\editors{Accepted to MIDL 2020}

\title[Bayesian QSM]{Bayesian Learning of Probabilistic Dipole Inversion for Quantitative Susceptibility Mapping}

\midlauthor{\Name{Jinwei Zhang\nametag{$^{1,2}$}} \Email{jz853@cornell.edu}\\
\addr $^{1}$ Department of Biomedical Engineering, Cornell University, Ithaca, NY, USA \\
\addr $^{2}$ Department of Radiology, Weill Medical College of Cornell University, New York, NY, USA \AND
\Name{Hang Zhang\midlotherjointauthor\nametag{$^{2,3}$}} \Email{hz459@cornell.edu}\\
\Name{Mert Sabuncu\nametag{$^{1,2,3}$}} \Email{msabuncu@cornell.edu}\\
\addr $^{3}$ Department of Electrical and Computer Engineering, Cornell University, Ithaca, NY, USA \AND
\Name{Pascal Spincemaille\midlotherjointauthor\nametag{$^{2}$}} \Email{pas2018@med.cornell.edu}\\
\Name{Thanh Nguyen\nametag{$^{2}$}} \Email{tdn2001@med.cornell.edu}\\
\Name{Yi Wang\nametag{$^{1,2}$}} \Email{yiwang@med.cornell.edu}\\
}

\begin{document}

\maketitle

\begin{abstract}
A learning-based posterior distribution estimation method, Probabilistic Dipole Inversion (PDI), is proposed to solve quantitative susceptibility mapping (QSM) inverse problem in MRI with uncertainty estimation. A deep convolutional neural network (CNN) is used to represent the multivariate Gaussian distribution as the approximated posterior distribution of susceptibility given the input measured field. In PDI, such CNN is firstly trained on healthy subjects' data with labels by maximizing the posterior Gaussian distribution loss function as used in Bayesian deep learning. When tested on new dataset without any label, PDI updates the pre-trained CNN's weights in an unsupervised fashion by minimizing the \emph{Kullback–Leibler} divergence between the approximated posterior distribution represented by CNN and the true posterior distribution given the likelihood distribution from known physical model and pre-defined prior distribution. Based on our experiments, PDI provides additional uncertainty estimation compared to the conventional MAP approach, meanwhile addressing the potential discrepancy issue of CNN when test data deviates from training dataset.          
\end{abstract}


\section{Introduction}

Consider the following biomedical imaging model:
\begin{align}
    y = Ax + n
\end{align}
where $A$ the forward imaging system model, $x$ the underlining biomedical image variable, $n$ the system noise, and $y$ the measured data/signal variable. Because of the intrinsic ill-posedness of forward imaging operator $A$, prior term is needed in the following Maximum a posteriori (MAP) estimation problem \cite{kaipio2006statistical}:

\begin{align}
    \hat{x} = \arg \max_x p(x|y) \propto p(y|x)p(x)    
\end{align}
where $p(x)$ is the prior term to regularize the inverse problem. Assuming zero mean Gaussian noise with covariance matrix $\Sigma$, Eq. 2 is equivalent to the following minimum $-\log p(x|y)$ problem:

\begin{align}
    \hat{x} = \arg \min_x || Ax - y ||^2_{\Sigma^{-1/2}} + R(x)
\end{align}
where $R(x) = -\log p(x)$. Convex optimization solvers have been widely used to solve Eq. 3 with both accuracy and efficiency, such as quasi-newton method \cite{dennis1977quasi}, alternating direction method of multipliers (ADMM) \cite{boyd2011distributed} and primal dual method (PD) \cite{chambolle2011first}.

In recent years, posterior distribution estimation in imaging inverse problems has been a new topic in medical imaging field \cite{repetti2019scalable, chappell2009variational, tezcan2018mr}, in which random variable's variance is provided from posterior distribution to measure the uncertainty of the solution. However, posterior distribution estimation requires complicated or even intractable integral from Bayes formula, therefore sampling or approximation method is used to reduce the computational cost and intractability of the problem. Markov chain Monte Carlo (MCMC) \cite{andrieu2003introduction} and variational inference (VI) \cite{bishop2006pattern} are two common frameworks in Bayesian estimation problem. In MCMC, efficient sampling methods are used to get random samples from posterior distribution. After proper sampling procedure, random samples can represent an empirical distribution which resembles the true distribution. However, in imaging inverse problem, the computational cost of approximating integrals for Bayesian estimation is often several magnitude higher than the optimization method of MAP estimation, suffering from curse of dimensionality \cite{pereyra2017maximum}. 

An alternative approach is to use VI, in which an approximation distribution is proposed with specific function form and unknown parameters, and then optimization algorithm is implemented (for example, expectation-maximization (EM) algorithm \cite{blei2017variational}) to learn these parameters by minimizing the divergence between true posterior and approximate posterior. After learning/fitting, the approximate posterior represents the true posterior. However, approximation quality is determined by the flexibility of approximate function form and trainable parameters. More complicated approximate function has better representation ability, however, the computational cost becomes higher. Therefore, the trade-off between number of trainable parameters of approximate function and learning/fitting efficiency needs careful consideration for VI.

Over the past years, thanks to the advances of deep learning, using deep neural network as the approximate function has become a new trend in VI, especially for generative models \cite{rezende2014stochastic, kingma2013auto}, in which low dimensional latent space variables are modeled and encoder and decoder networks are built to approximate the latent variable distribution conditioned on observed data and reversely observed data distribution conditioned on latent variable. Due to the approximation and generalization power of deep neural network with millions of trainable weights, neural network can approximate any function/distribution with high accuracy. In addition, advanced stochastic optimization algorithms such as ADAM \cite{kingma2014adam} have been proposed for efficient backpropagation in network weights' updating. Another topic related to the posterior distribution estimation with deep learning is discussed in Bayesian deep learning framework \cite{kendall2017uncertainties}, in which data uncertainties are captured by maximizing the posterior distribution with the labels as the samples assuming they follow multivariate Gaussian. 

In this paper, we come up with a framework by combining Bayesian deep learning to model data uncertainties and VI with deep learning to approximate true posterior distribution, and apply it to one important imaging inverse problem in MRI: \emph{quantitative susceptibility mapping} (QSM) \cite{de2010quantitative, wang2015quantitative}, which has the advantages of mapping iron decomposition \cite{wang2017clinical} and calcification \cite{chen2014intracranial}. Assuming multivariate Gaussian represented by a CNN as the posterior distribution of susceptibility given the input local field, golden standard susceptibility maps COSMOS (Calculation Of Susceptibility through Multiple Orientation Sampling \cite{liu2009calculation}) are used to train such CNN with a maximal posterior loss function.  With physical model-based likelihood term and delicately designed prior term, the pre-trained CNN can be enhanced when tested on patient dataset by minimizing the Kullback-Leibler (KL) divergence between true posterior distribution and approximation distribution represented by such CNN. Our experimental results show the proposed method gives mean and variance estimation of the solution automatically, and yields optimal results compared to two types of benchmark methods: deep learning QSM (\cite{yoon2018quantitative, zhang2020fidelity}) and \emph{maximum a posteriori} (MAP) QSM with convex optimization \cite{liu2012morphology, kee2017quantitative, milovic2018fast}, both of which do not provide uncertainty estimation.

\section{Method}

\subsection{Modeling}
In Magnetic Resonance Imaging (MRI), the forward model of generating relative local field $b$ from tissue susceptibility $\chi$ is:
\begin{align}
   b = d \ast \chi + n 
\end{align}
where $\ast$ denotes convolution operation, $d$ denotes \textit{dipole kernel}, which is ill-posed inherent in the structure of dipole convolution operator. $b$ is derived from multi-echo gradient echo MR signal with noise $n$ estimated as well. The inverse problem of estimating $\chi$ from measured $b$ is called \textit{Quantitative Susceptibility Mapping} (QSM). From convolution theory, the forward convolution process in Eq. 4 is equivalent to the following Fourier space multiplication process:
\begin{align}
   b = F^HDF\chi + n 
\end{align}
where $F$ is Fourier matrix, $D$ is dipole kernel in Fourier space. Eq. 5 is computationally friendly since Fast Fourier Transform (FFT) can be used efficiently. We will use forward model in Eq. 5 for computation in this paper.

One successful approach to solving QSM from single orientation field $b$ is MEDI (Morphology enabled dipole inversion) \cite{liu2011morphology, liu2012morphology}, where weighted total variation regularization was imposed onto the area except tissue in brain and the following MAP estimation is deployed:

\begin{align}
    \hat{\chi} = \arg\min_{\chi}|| W(F^HDF\chi - b) ||^2_2 + \lambda|| M\nabla \chi||_1
\end{align}
where $W$ is derived from observation noise covariance matrix and $M$ is gradient's weight to penalize only region's outside brain tissues. Computational methods for solving Eq. 6 is reviewed in \cite{kee2017quantitative}.

Starting from forward model in Eq. 5, we develop the fully probabilistic model of QSM and use approximate Bayesian inference to solve this problem. We assume conditional distribution of field $b$ given susceptibility $\chi$ as a Gaussian distribution:
\begin{align}
    p(b|\chi) = \mathcal{N}(b|F^HDF\chi, \Sigma_{b|\chi})
\end{align}
where we assume $n \sim \mathcal{N}(0,\Sigma_{b|\chi})$ with $\Sigma_{b|\chi}$ diagonal in Eq. 5. The prior distribution from Eq. 6 reads:
\begin{align}
    p(\chi) \propto {\rm e}^{-\lambda\| M\nabla \chi \|_1}.
\end{align}
Other types of prior distributions can also be applied. Because of the intractability of estimating the posterior distribution $p(\chi|b) = p(b|\chi)p(\chi)/ \int_{\chi} p(b|\chi)p(\chi) d\chi$ in most cases, approximate posterior distribution $q(\chi|b) = \mathcal{N}(\mu_{\chi|b},\Sigma_{\chi|b})$ with diagonal covariance matrix is assumed to approximate the true posterior distribution $p(\chi|b)$. In this work, we use a dual-decoder network architecture (Figure 1) extended from 3D U-Net \cite{ronneberger2015u, cciccek20163d} to represent the approximate posterior $q_{\psi}(\chi|b)$, with each decoder's output representing mean $\mu_{\chi|b}$ and variance $\Sigma_{\chi|b}$ map, respectively.

\subsection{Supervised Bayesian Training}
For training dataset with COSMOS as golden standard labels, we can treat these labels as samples from the true posterior distribution, and train the approximate distribution $q_{\psi}(\chi|b)$ in a supervised fashion with the following MAP loss function:
\begin{align}
    - \log q_{\psi}(\chi_i|b_i) = \frac{1}{2}(\chi_i - \mu_{\chi|b_i})^T\Sigma_{\chi|b_i}^{-1}(\chi_i - \mu_{\chi|b_i}) + \frac{1}{2} \ln |\Sigma_{\chi|b_i}|,
\end{align}
where $\{b_i, \chi_i\}$ denote the i-th input and label data pair in the training dataset. Note that because of multiple orientations' scanning for COSMOS, this dataset is quite limited and usually only on healthy subjects. We denote this supervised Bayesian learning approach as Probabilistic Dipole Inversion (PDI).

\begin{figure}[t]
 % Caption and label go in the first argument and the figure contents
 % go in the second argument
\floatconts
  {fig:fig1}
  {\caption{Network architecture of the proposed method.  Dual decoders' outputs represent mean and variance maps. COSMOS dataset was used to do supervised Bayesian training via MAP in Eq. 9. Unsupervised VI with MC sampling in Eq. 11 and 12 was applied on other test dataset.}}
  {\includegraphics[width=0.9\textwidth]{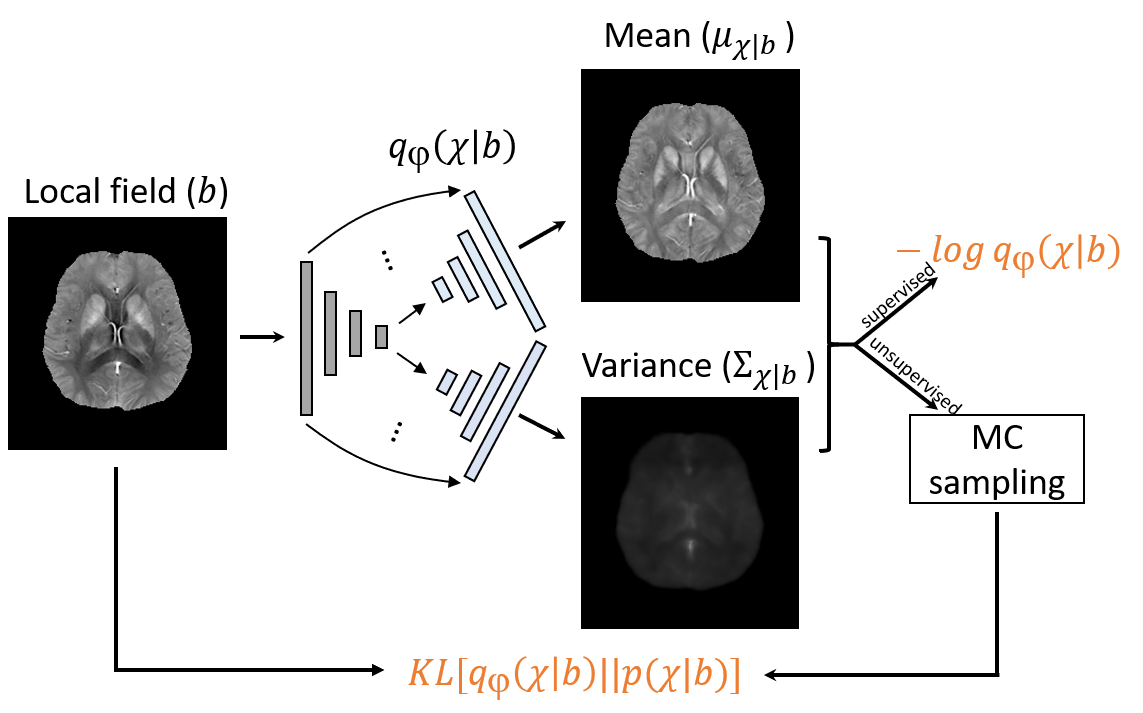}}
\end{figure}

\subsection{Unsupervised Variational Inference}
After training using COSMOS data with loss function Eq. 9 and obtaining optimal parameters $\psi^*$, given a test local field $b^{'}$, we can simply estimate $p(\chi|b^{'})$ as $q_{\psi^*}(\chi|b^{'})$. However, for new test dataset which has input field $b^{'}$ deviating from COSMOS training dataset (such as having new pathologies), inferior outputs could be produced. In this case, $q_{\psi^*}(\chi|b^{'})$ can be enhanced by deploying variational inference on a subset of this new test dataset as another training set. specifically, the pre-trained approximation $q_{\psi}(\chi|b^{'})$ with weights $\psi$ initialized as $\psi^*$ can be fine-tuned by minimizing the KL divergence between $p(\chi|b^{'})$ and $q_{\psi}(\chi|b^{'})$:
\begin{align}
\begin{split}
    & \text{ KL}[q_{\psi}(\chi|b^{'})||p(\chi|b^{'})]\\
     = \ &\mathbb{E}_q[\log q_{\psi}(\chi|b^{'}) - \log p(\chi|b^{'})]\\
     = \ &\mathbb{E}_q[\log q_{\psi}(\chi|b^{'}) - \log p(\chi,b^{'})] + \log p(b^{'})\\\
     = \ &\text{KL}[q_{\psi}(\chi|b^{'})||p(\chi)] - \mathbb{E}_q[\log p(b^{'}|\chi)]
\end{split}
\end{align}
where the first term in the last equation above imposes the posterior to be similar to the prior, and the second term encourages data consistency in QSM foward model. Applying the prior term defined in Eq. 8 and likelihood term in Eq. 7, KL divergence in Eq. 10 becomes:
\begin{align}
\begin{split}
     &\text{ KL}[q_{\psi}(\chi|b^{'})||p(\chi|b^{'})]\\
    = \ & - \frac{1}{2}\text{ln}|\Sigma_{\chi|b^{'}}|  + \frac{1}{2K}\sum_{k=1}^{K}\lambda \| M\nabla \chi_k\|_1 + \frac{1}{2K}\sum_{k=1}^{K}(\chi_k \ast d - b^{'})^T\Sigma_{b^{'}|\chi}^{-1}(\chi_k \ast d - b^{'}) 
\end{split}
\end{align}
where $- \mathbb{E}_q[\ln p(\chi)]$ in $\text{KL}[q_\psi(\chi|b^{'})||p(\chi)]$ and $- \mathbb{E}_q[\log p(b^{'}|\chi)]$ are calculated through Monle Carlo (MC) sampling with $\chi_k$ sampled from $q_{\psi}(\chi|b^{'})$ because of the  intractability of both expectations. We denote the fine-tuned approximate distribution with Eq. 11 as PDI-VI1.

Another possible prior term for $\chi$ is simply constant prior $p(\chi) \propto c$, which means no prior information is given regarding the distribution of $\chi$. With such non-informative prior, the corresponding loss function is simply:
\begin{align}
\begin{split}
     \text{ KL}[q_{\psi}(\chi|b^{'})||p(\chi|b^{'})]
    = \ - \frac{1}{2}\text{ln}|\Sigma_{\chi|b^{'}}| + \frac{1}{2K}\sum_{k=1}^{K}(\chi_k \ast d - b^{'})^T\Sigma_{b^{'}|\chi}^{-1}(\chi_k \ast d - b^{'}) 
\end{split}
\end{align}
We denote the fine-tuned approximate distribution with Eq. 12 as PDI-VI2.

\section{Experiments}

\begin{figure*}[t]
  \centering
  \includegraphics[width=1\textwidth]{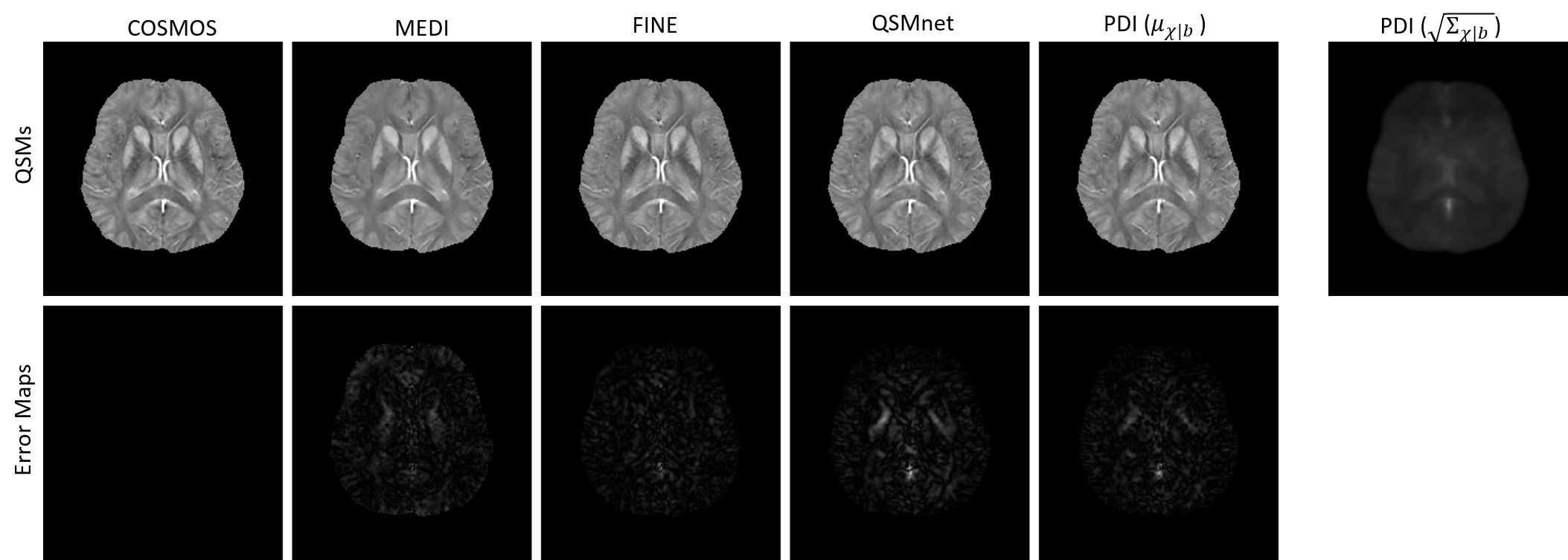}
  \caption{Reconstructions (first row) and error maps (second row) of one COSMOS test subject in one orientation, with COSMOS as the golden standard. FINE gives the best reconstruction at the expense of significantly increased computational time. The other three methods have comparable results. The standard deviation map (last column) provided by PDI resembled its error map, with high uncertainties/errors locating at sagittal sinus and globus pallidus.}
\end{figure*}

MRI was performed on 7 healthy subjects with 5 brain orientations using a 3T GE scanner equipped with a multi-echo 3D gradient echo (GRE) sequence. Acquisition matrix was $256\times256\times48$ and voxel size was $1\times1\times3 \ \text{mm}^3$. The input local tissue field data $b$ was generated by sequentially deploying non-linear fitting across multi-echo phase data \cite{kressler2009nonlinear}, graph-cut based phase unwrapping \cite{dong2014simultaneous} and background field removal \cite{liu2011novel}. COSMOS reconstruction \cite{liu2011morphology} was calculated from 5 orientations' GRE imaging and was used as the gold standard label in the experiment. A second dataset was obtained by performing single orientation GRE MRI on 8 patients with intracerebral hemorrhage (ICH), which were acquired using the same scanner and imaging parameters as above.

\begin{figure*}[t]
  \centering
  \includegraphics[width=1\textwidth]{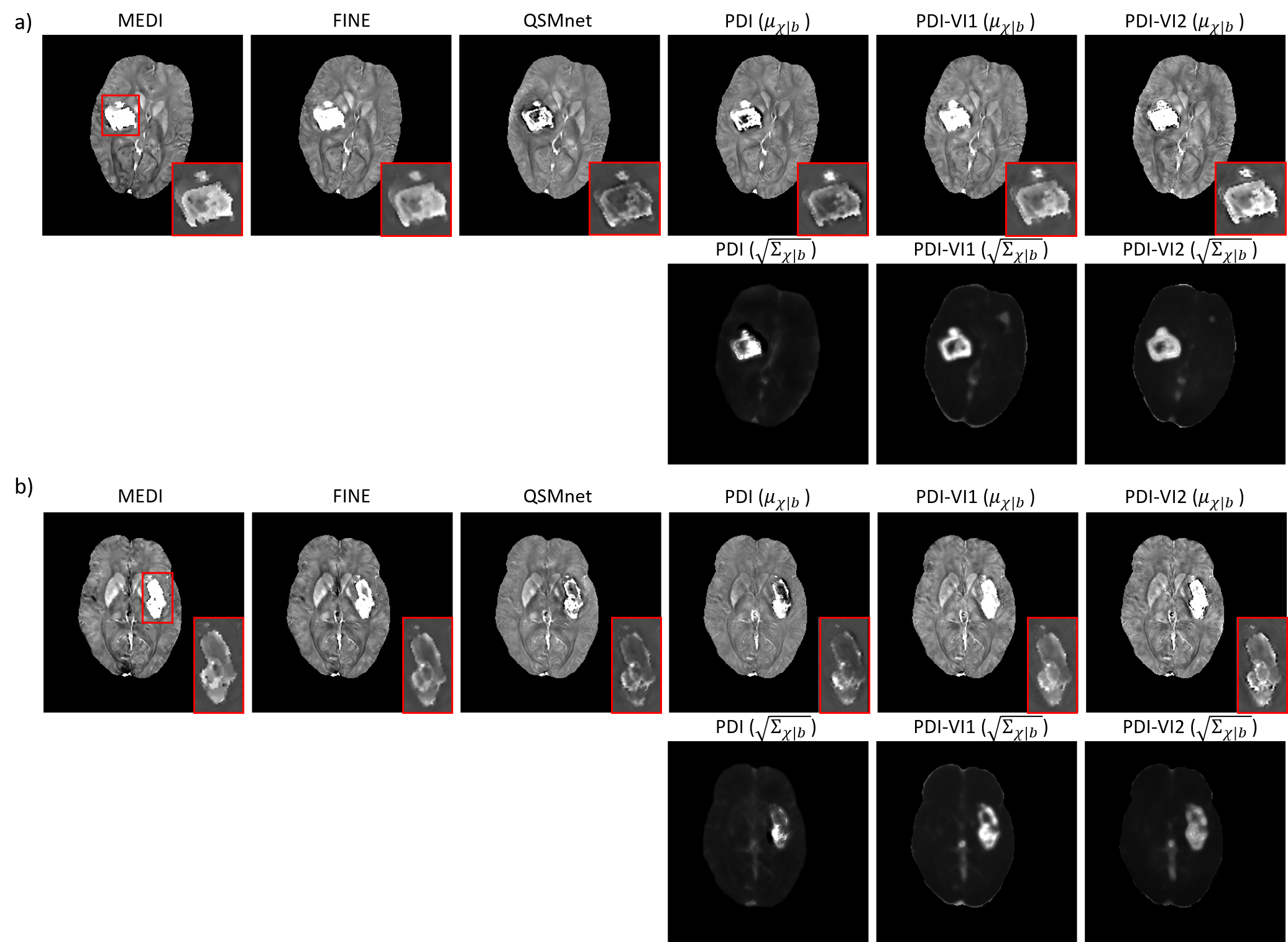}
  \caption{Reconstructions (first row in (a) and (b)) and standard deviation maps (second row in (a) and (b)) of two ICH patients. Compared to MEDI and FINE, underestimation issue inside hemorrhage happened on QSMnet and PDI. This issue was reduced in PDI-VI1 and PDI-VI2 by fine-tuning the pre-trained network using unsupervised variational inference. High variance inside the hemorrhage was consistent with high measured noise in the same region.}
\end{figure*}

Network architecture is shown in Figure 1. Dual decoders' outputs were used to represent mean and variance maps in the posterior susceptibility distribution given input local field. The 3D convolutional kernel size was $3\times3\times3$. The numbers of filters from the highest feature level to the lowest were 32, 64, 128, 256 and 512, respectively. Batch normalization \cite{ioffe2015batch}, max pooling for downsampling and deconvolution operation for upsampling were used. For COSMOS dataset, 4/1 subjects (20/5 brain volumes) were used as training/validation dataset, with augmentation by in-plane rotation of $\pm15^{\circ}$. Each brain volume data in the training and validation dataset was divided into 3D patches with patch size $64\times64\times32$ and extraction step $21\times21\times11$. The remaining 2 subjects (10 brain volumes in total) were used for testing. For ICH patients dataset, 5/1 subjects were used as training/validation dataset for PDI-VI1 and PDI-VI, and the remaining 2 subjects were used for testing. 

Loss function in Eq. 9 was applied for supervised Bayesian training on COSMOS dataset with ADAM optimizer \cite{kingma2014adam} (learning rate: $10^{-3}$, Number of epochs: 60), yielding trained network $q_{\psi^*}(\chi|b)$. The outputs of $q_{\psi^*}(\chi|b)$ were denoted as PDI. Initialized with the pre-trained PDI using COSMOS training data, unsupervised variational inference with loss function Eq. 11 and 12 was also applied on ICH dataset using ADAM optimizer (learning rate: $10^{-3}$, Number of iterations: 100). MC sampling size $K$ was chosen as $5$ due to limited GPU memory and reparameterization trick \cite{kingma2013auto} was used for MC sampling in order to do backpropagation. The outputs were denoted as PDI-VI1 and PDI-VI2, respectively. The whole brain volume was fed into the network during testing, including unsupervised variational inference step. We implemented the proposed method using PyTorch (Python 3.6) on an RTX 2080Ti GPU.

\begin{table}
\captionsetup{font=small}
\begin{tabular}{cccccc} \hline
   & pSNR & RMSE & SSIM & HFEN & GPU time (s) \\ \hline
MEDI \cite{liu2012morphology} & 46.39 & 41.16 & 0.9569 & 31.30 & 17.54\\
FINE \cite{zhang2020fidelity} & 48.12 & 33.66 & 0.9789 & 31.97 & 65.42\\
QSMnet \cite{yoon2018quantitative} & 46.35 & 41.29 & 0.9705 & 43.31 & 0.60\\ 
PDI (Eq. 9) & 47.77 & 35.08 & 0.9772 & 35.17 & 0.61\\
\hline
\\
\end{tabular}
\caption{Mean quantitative metrics of 10 test COSMOS brains reconstructed by different methods. FINE gives the best reconstruction at the expense of significantly increased computational time.  The other three methods have comparable results.}
\end{table}

For COSMOS test dataset, we compared PDI with MAP estimation MEDI \cite{liu2012morphology} and two deep learning reconstructions QSMnet \cite{yoon2018quantitative} and FINE \cite{zhang2020fidelity}. Reconstruction maps of one orientation from one test subject are shown in Figure 2 ([-0.15ppm, 0.15ppm]). Quantitative metrics of each reconstruction method averaged among 10 test brains are shown in Table 1. FINE gave the best overall quantitative results; However, it overfitted to every test case by minimizing the fidelity loss, which had the major drawback of significantly increased computational time. PDI gave slightly better results than MEDI and QSMnet, meanwhile achieved fast inference time on GPU comparable to QSMnet. In figure 2, error map of PDI's mean output $\mu_{\chi|b}$ was coincident with PDI's standard deviation output $\sqrt{\Sigma_{\chi|b}}$, with high uncertainty/error happening at sagittal sinus and globus pallidus.

For ICH test dataset, PDI-VI1 and PDI-VI2 were also performed and compared. Two representative ICH patients' QSMs are shown in Figure 3 ([-0.6ppm, 1.5ppm] for zoomed-in hemorrhage). Compared to MEDI and FINE which had hyperintensity inside the hemorrhage, both QSMnet and PDI suffered from underestimation issue inside this region, which might result from the fact that such pathology was not encountered during training since long scan COSMOS was not practical for the patients. After PDI-VI1 and PDI-VI2, such underestimation issue was reduced and variance maps' structures inside the hemorrhage were also better depicted. High uncertainties inside hemorrhage as shown in Figure 3 were consistent with high local field noise level which was approximately proportional to the underlining susceptibility values.

\section{Conclusion}
We developed a Bayesian dipole inversion framework for quantitative susceptibility mapping by combining variational inference and Bayesian deep learning. Our method generated high fidelity susceptibility maps meanwhile provided uncertainty quantifications. When applied to other datasets not encountered during training, the proposed method was able to correct the undesirable outputs in an unsupervised fashion based on variantional inference principle.

% Acknowledgments---Will not appear in anonymized version
% \midlacknowledgments{We thank a bunch of people.}

\bibliography{midl-samplebibliography}

\begin{thebibliography}{32}
\providecommand{\natexlab}[1]{#1}
\providecommand{\url}[1]{\texttt{#1}}
\expandafter\ifx\csname urlstyle\endcsname\relax
  \providecommand{\doi}[1]{doi: #1}\else
  \providecommand{\doi}{doi: \begingroup \urlstyle{rm}\Url}\fi

\bibitem[Andrieu et~al.(2003)Andrieu, De~Freitas, Doucet, and
  Jordan]{andrieu2003introduction}
Christophe Andrieu, Nando De~Freitas, Arnaud Doucet, and Michael~I Jordan.
\newblock An introduction to mcmc for machine learning.
\newblock \emph{Machine learning}, 50\penalty0 (1-2):\penalty0 5--43, 2003.

\bibitem[Bishop(2006)]{bishop2006pattern}
Christopher~M Bishop.
\newblock \emph{Pattern recognition and machine learning}.
\newblock springer, 2006.

\bibitem[Blei et~al.(2017)Blei, Kucukelbir, and McAuliffe]{blei2017variational}
David~M Blei, Alp Kucukelbir, and Jon~D McAuliffe.
\newblock Variational inference: A review for statisticians.
\newblock \emph{Journal of the American Statistical Association}, 112\penalty0
  (518):\penalty0 859--877, 2017.

\bibitem[Boyd et~al.(2011)Boyd, Parikh, Chu, Peleato, Eckstein,
  et~al.]{boyd2011distributed}
Stephen Boyd, Neal Parikh, Eric Chu, Borja Peleato, Jonathan Eckstein, et~al.
\newblock Distributed optimization and statistical learning via the alternating
  direction method of multipliers.
\newblock \emph{Foundations and Trends{\textregistered} in Machine learning},
  3\penalty0 (1):\penalty0 1--122, 2011.

\bibitem[Chambolle and Pock(2011)]{chambolle2011first}
Antonin Chambolle and Thomas Pock.
\newblock A first-order primal-dual algorithm for convex problems with
  applications to imaging.
\newblock \emph{Journal of mathematical imaging and vision}, 40\penalty0
  (1):\penalty0 120--145, 2011.

\bibitem[Chappell et~al.(2009)Chappell, Groves, Whitcher, and
  Woolrich]{chappell2009variational}
Michael~A Chappell, Adrian~R Groves, Brandon Whitcher, and Mark~W Woolrich.
\newblock Variational bayesian inference for a nonlinear forward model.
\newblock \emph{IEEE Transactions on Signal Processing}, 57\penalty0
  (1):\penalty0 223--236, 2009.

\bibitem[Chen et~al.(2014)Chen, Zhu, Kovanlikaya, Kovanlikaya, Liu, Wang,
  Salustri, and Wang]{chen2014intracranial}
Weiwei Chen, Wenzhen Zhu, IIhami Kovanlikaya, Arzu Kovanlikaya, Tian Liu, Shuai
  Wang, Carlo Salustri, and Yi~Wang.
\newblock Intracranial calcifications and hemorrhages: characterization with
  quantitative susceptibility mapping.
\newblock \emph{Radiology}, 270\penalty0 (2):\penalty0 496--505, 2014.

\bibitem[{\c{C}}i{\c{c}}ek et~al.(2016){\c{C}}i{\c{c}}ek, Abdulkadir, Lienkamp,
  Brox, and Ronneberger]{cciccek20163d}
{\"O}zg{\"u}n {\c{C}}i{\c{c}}ek, Ahmed Abdulkadir, Soeren~S Lienkamp, Thomas
  Brox, and Olaf Ronneberger.
\newblock 3d u-net: learning dense volumetric segmentation from sparse
  annotation.
\newblock In \emph{International conference on medical image computing and
  computer-assisted intervention}, pages 424--432. Springer, 2016.

\bibitem[de~Rochefort et~al.(2010)de~Rochefort, Liu, Kressler, Liu,
  Spincemaille, Lebon, Wu, and Wang]{de2010quantitative}
Ludovic de~Rochefort, Tian Liu, Bryan Kressler, Jing Liu, Pascal Spincemaille,
  Vincent Lebon, Jianlin Wu, and Yi~Wang.
\newblock Quantitative susceptibility map reconstruction from mr phase data
  using bayesian regularization: validation and application to brain imaging.
\newblock \emph{Magnetic Resonance in Medicine: An Official Journal of the
  International Society for Magnetic Resonance in Medicine}, 63\penalty0
  (1):\penalty0 194--206, 2010.

\bibitem[Dennis and Mor{\'e}(1977)]{dennis1977quasi}
John~E Dennis, Jr and Jorge~J Mor{\'e}.
\newblock Quasi-newton methods, motivation and theory.
\newblock \emph{SIAM review}, 19\penalty0 (1):\penalty0 46--89, 1977.

\bibitem[Dong et~al.(2014)Dong, Liu, Chen, Zhou, Dimov, Raj, Cheng,
  Spincemaille, and Wang]{dong2014simultaneous}
Jianwu Dong, Tian Liu, Feng Chen, Dong Zhou, Alexey Dimov, Ashish Raj, Qiang
  Cheng, Pascal Spincemaille, and Yi~Wang.
\newblock Simultaneous phase unwrapping and removal of chemical shift (spurs)
  using graph cuts: application in quantitative susceptibility mapping.
\newblock \emph{IEEE transactions on medical imaging}, 34\penalty0
  (2):\penalty0 531--540, 2014.

\bibitem[Ioffe and Szegedy(2015)]{ioffe2015batch}
Sergey Ioffe and Christian Szegedy.
\newblock Batch normalization: Accelerating deep network training by reducing
  internal covariate shift.
\newblock \emph{arXiv preprint arXiv:1502.03167}, 2015.

\bibitem[Kaipio and Somersalo(2006)]{kaipio2006statistical}
Jari Kaipio and Erkki Somersalo.
\newblock \emph{Statistical and computational inverse problems}, volume 160.
\newblock Springer Science \& Business Media, 2006.

\bibitem[Kee et~al.(2017)Kee, Liu, Zhou, Dimov, Cho, De~Rochefort, Seo, and
  Wang]{kee2017quantitative}
Youngwook Kee, Zhe Liu, Liangdong Zhou, Alexey Dimov, Junghun Cho, Ludovic
  De~Rochefort, Jin~Keun Seo, and Yi~Wang.
\newblock Quantitative susceptibility mapping (qsm) algorithms: mathematical
  rationale and computational implementations.
\newblock \emph{IEEE Transactions on Biomedical Engineering}, 64\penalty0
  (11):\penalty0 2531--2545, 2017.

\bibitem[Kendall and Gal(2017)]{kendall2017uncertainties}
Alex Kendall and Yarin Gal.
\newblock What uncertainties do we need in bayesian deep learning for computer
  vision?
\newblock In \emph{Advances in neural information processing systems}, pages
  5574--5584, 2017.

\bibitem[Kingma and Ba(2014)]{kingma2014adam}
Diederik~P Kingma and Jimmy Ba.
\newblock Adam: A method for stochastic optimization.
\newblock \emph{arXiv preprint arXiv:1412.6980}, 2014.

\bibitem[Kingma and Welling(2013)]{kingma2013auto}
Diederik~P Kingma and Max Welling.
\newblock Auto-encoding variational bayes.
\newblock \emph{arXiv preprint arXiv:1312.6114}, 2013.

\bibitem[Kressler et~al.(2009)Kressler, De~Rochefort, Liu, Spincemaille, Jiang,
  and Wang]{kressler2009nonlinear}
Bryan Kressler, Ludovic De~Rochefort, Tian Liu, Pascal Spincemaille, Quan
  Jiang, and Yi~Wang.
\newblock Nonlinear regularization for per voxel estimation of magnetic
  susceptibility distributions from mri field maps.
\newblock \emph{IEEE transactions on medical imaging}, 29\penalty0
  (2):\penalty0 273--281, 2009.

\bibitem[Liu et~al.(2012)Liu, Liu, de~Rochefort, Ledoux, Khalidov, Chen,
  Tsiouris, Wisnieff, Spincemaille, Prince, et~al.]{liu2012morphology}
Jing Liu, Tian Liu, Ludovic de~Rochefort, James Ledoux, Ildar Khalidov, Weiwei
  Chen, A~John Tsiouris, Cynthia Wisnieff, Pascal Spincemaille, Martin~R
  Prince, et~al.
\newblock Morphology enabled dipole inversion for quantitative susceptibility
  mapping using structural consistency between the magnitude image and the
  susceptibility map.
\newblock \emph{Neuroimage}, 59\penalty0 (3):\penalty0 2560--2568, 2012.

\bibitem[Liu et~al.(2009)Liu, Spincemaille, De~Rochefort, Kressler, and
  Wang]{liu2009calculation}
Tian Liu, Pascal Spincemaille, Ludovic De~Rochefort, Bryan Kressler, and
  Yi~Wang.
\newblock Calculation of susceptibility through multiple orientation sampling
  (cosmos): a method for conditioning the inverse problem from measured
  magnetic field map to susceptibility source image in mri.
\newblock \emph{Magnetic Resonance in Medicine: An Official Journal of the
  International Society for Magnetic Resonance in Medicine}, 61\penalty0
  (1):\penalty0 196--204, 2009.

\bibitem[Liu et~al.(2011{\natexlab{a}})Liu, Khalidov, de~Rochefort,
  Spincemaille, Liu, Tsiouris, and Wang]{liu2011novel}
Tian Liu, Ildar Khalidov, Ludovic de~Rochefort, Pascal Spincemaille, Jing Liu,
  A~John Tsiouris, and Yi~Wang.
\newblock A novel background field removal method for mri using projection onto
  dipole fields.
\newblock \emph{NMR in Biomedicine}, 24\penalty0 (9):\penalty0 1129--1136,
  2011{\natexlab{a}}.

\bibitem[Liu et~al.(2011{\natexlab{b}})Liu, Liu, De~Rochefort, Spincemaille,
  Khalidov, Ledoux, and Wang]{liu2011morphology}
Tian Liu, Jing Liu, Ludovic De~Rochefort, Pascal Spincemaille, Ildar Khalidov,
  James~Robert Ledoux, and Yi~Wang.
\newblock Morphology enabled dipole inversion (medi) from a single-angle
  acquisition: comparison with cosmos in human brain imaging.
\newblock \emph{Magnetic resonance in medicine}, 66\penalty0 (3):\penalty0
  777--783, 2011{\natexlab{b}}.

\bibitem[Milovic et~al.(2018)Milovic, Bilgic, Zhao, Acosta-Cabronero, and
  Tejos]{milovic2018fast}
Carlos Milovic, Berkin Bilgic, Bo~Zhao, Julio Acosta-Cabronero, and Cristian
  Tejos.
\newblock Fast nonlinear susceptibility inversion with variational
  regularization.
\newblock \emph{Magnetic resonance in medicine}, 80\penalty0 (2):\penalty0
  814--821, 2018.

\bibitem[Pereyra(2017)]{pereyra2017maximum}
Marcelo Pereyra.
\newblock Maximum-a-posteriori estimation with bayesian confidence regions.
\newblock \emph{SIAM Journal on Imaging Sciences}, 10\penalty0 (1):\penalty0
  285--302, 2017.

\bibitem[Repetti et~al.(2019)Repetti, Pereyra, and Wiaux]{repetti2019scalable}
Audrey Repetti, Marcelo Pereyra, and Yves Wiaux.
\newblock Scalable bayesian uncertainty quantification in imaging inverse
  problems via convex optimization.
\newblock \emph{SIAM Journal on Imaging Sciences}, 12\penalty0 (1):\penalty0
  87--118, 2019.

\bibitem[Rezende et~al.(2014)Rezende, Mohamed, and
  Wierstra]{rezende2014stochastic}
Danilo~Jimenez Rezende, Shakir Mohamed, and Daan Wierstra.
\newblock Stochastic backpropagation and approximate inference in deep
  generative models.
\newblock \emph{arXiv preprint arXiv:1401.4082}, 2014.

\bibitem[Ronneberger et~al.(2015)Ronneberger, Fischer, and
  Brox]{ronneberger2015u}
Olaf Ronneberger, Philipp Fischer, and Thomas Brox.
\newblock U-net: Convolutional networks for biomedical image segmentation.
\newblock In \emph{International Conference on Medical image computing and
  computer-assisted intervention}, pages 234--241. Springer, 2015.

\bibitem[Tezcan et~al.(2018)Tezcan, Baumgartner, Luechinger, Pruessmann, and
  Konukoglu]{tezcan2018mr}
Kerem~C Tezcan, Christian~F Baumgartner, Roger Luechinger, Klaas~P Pruessmann,
  and Ender Konukoglu.
\newblock Mr image reconstruction using deep density priors.
\newblock \emph{IEEE transactions on medical imaging}, 2018.

\bibitem[Wang and Liu(2015)]{wang2015quantitative}
Yi~Wang and Tian Liu.
\newblock Quantitative susceptibility mapping (qsm): decoding mri data for a
  tissue magnetic biomarker.
\newblock \emph{Magnetic resonance in medicine}, 73\penalty0 (1):\penalty0
  82--101, 2015.

\bibitem[Wang et~al.(2017)Wang, Spincemaille, Liu, Dimov, Deh, Li, Zhang, Yao,
  Gillen, Wilman, et~al.]{wang2017clinical}
Yi~Wang, Pascal Spincemaille, Zhe Liu, Alexey Dimov, Kofi Deh, Jianqi Li, Yan
  Zhang, Yihao Yao, Kelly~M Gillen, Alan~H Wilman, et~al.
\newblock Clinical quantitative susceptibility mapping (qsm): biometal imaging
  and its emerging roles in patient care.
\newblock \emph{Journal of Magnetic Resonance Imaging}, 46\penalty0
  (4):\penalty0 951--971, 2017.

\bibitem[Yoon et~al.(2018)Yoon, Gong, Chatnuntawech, Bilgic, Lee, Jung, Ko,
  Jung, Setsompop, Zaharchuk, et~al.]{yoon2018quantitative}
Jaeyeon Yoon, Enhao Gong, Itthi Chatnuntawech, Berkin Bilgic, Jingu Lee, Woojin
  Jung, Jingyu Ko, Hosan Jung, Kawin Setsompop, Greg Zaharchuk, et~al.
\newblock Quantitative susceptibility mapping using deep neural network:
  Qsmnet.
\newblock \emph{Neuroimage}, 179:\penalty0 199--206, 2018.

\bibitem[Zhang et~al.(2020)Zhang, Liu, Zhang, Zhang, Spincemaille, Nguyen,
  Sabuncu, and Wang]{zhang2020fidelity}
Jinwei Zhang, Zhe Liu, Shun Zhang, Hang Zhang, Pascal Spincemaille, Thanh~D
  Nguyen, Mert~R Sabuncu, and Yi~Wang.
\newblock Fidelity imposed network edit (fine) for solving ill-posed image
  reconstruction.
\newblock \emph{NeuroImage}, page 116579, 2020.

\end{thebibliography}

\end{document}